\documentclass[12pt]{iopart}
\usepackage{cite}
\usepackage{txfonts}

\usepackage{graphicx}% Include figure files
\usepackage{dcolumn}% Align table columns on decimal point
\usepackage{bm}% bold math
\usepackage[usenames,dvipsnames]{color}

\begin{document}

\title[]{Coherent quantum
dynamics in donor-bridge-acceptor systems: Beyond the hopping and super-exchange mechanisms}

\author{Seogjoo Jang,$^{*,1,2}$ Timothy C Berkelbach,$^2$ and David R Reichman$^2$}
\address{$^1$Department of Chemistry and Biochemistry, Queens College of the City
University of New York, 65-30 Kissena Boulevard, Flushing, New York, 11367}
\address{$^2$Department of Chemistry, Columbia University, 3000 Broadway, New York, New
York 10027}

%\date{\today}

\begin{abstract}
The population transfer dynamics of 
model donor-bridge-acceptor systems is studied by comparing
a recently developed polaron-transformed quantum master
equation (PQME) with the well-known Redfield and F\"{o}rster theories
of quantum transport.  We show that the PQME approach
reduces to these two theories in their respective limits of validity
and naturally interpolates between them as a function of the system-bath
coupling strength.
By exploring the parameter space of our model problem, we identify novel
regimes of transport dynamics in bridged systems like those encountered
in biological and organic energy transfer problems.
Furthermore, we demonstrate that three-level systems
like the ones studied herein represent ideal minimal models for the
identification of quantum coherent transport as embodied in super-exchange
phenomena that cannot be captured by F\"{o}rster-like hopping approaches.
\end{abstract}

%\pacs{Valid PACS appear here}% PACS, the Physics and Astronomy Classification Scheme.
%\keywords{Suggested keywords}%Use showkeys class option if keyword display desired

\maketitle

\small

\section{Introduction}

The dynamics of charge and exciton migration in many
biological~\cite{adv-et,renger-pr343,yu-prl86,wang-science316,engel-nature466,collini-nature463}
and synthetic molecular
systems~\cite{adv-et,davis-nature396,thompson-jacs127,collini-science323,vura-weis-science328,lloveras_jacs133}
proceed through multiple sites or states embedded in a complex interacting environment.
Thus an accurate understanding of such processes has significant implications for
elucidating the microscopic mechanism of a variety of processes in chemistry, biology, and
materials science. Unfortunately, a quantitative description of electron and exciton
transfer dynamics in condensed phases is usually beyond the reach of present day
analytical and numerical techniques.  In particular, powerful exact numerical approaches
are generally limited to a small number of discrete quantum states and particular forms of
the environment and its interaction with those 
states\cite{mak92,egg94}. Approximate
analytical approaches are often tied to particular regions of parameter space where at
least one energy scale can be quantified as ``small" compared to all others.
Approximate numerical techniques, such as semiclassical
``surface hopping"~\cite{pec69,tul98} and mean-field~\cite{tul98,sto95}
approaches, have the ability to model large systems with more general reservoirs in a
computationally facile manner, but often fail in particular parameter regimes as well,
especially when quantum effects are important.
Furthermore, the range of validity of such techniques is often difficult to quantify.
Given this state of affairs, the development of approximate methods that can bridge the
gap between different parameter regimes accurately and that can be adapted to the
description of relatively large scale systems is of prime importance.

In terms of the methods mentioned above, quantum master equation (QME) approaches are
attractive because they offer the possibility of controlled approximations (that can be
corrected via higher order expansions), the ability to treat large numbers of quantum
states, and the flexibility to be paired with semiclassical approaches in the treatment of
bath correlations.  Standard expansion schemes make use of an approximation exact up to
the second order of either the system-bath coupling (Redfield theory and related
approaches~\cite{blo57,red65,bre02})
or the electronic coupling (the ``non-interacting blip
approximation"~\cite{leg87} yielding
Marcus~\cite{mar64} and F\"{o}rster~\cite{for53} theories for electron and energy transfer, respectively).
While these approaches are
successful when their respective expansion parameters are small, they can be grossly
inaccurate when extended beyond their limits of applicability.   An important recent
example of the distinction between these two limits comes from studies of energy
conversion processes such as singlet fission in organic
assemblies~\cite{berkelbach-jcp138-1,berkelbach-jcp138-2}. Higher order QMEs or the recently
developed polaronic QME (PQME) approaches~\cite{jang-jcp129,jang-jcp131,jang-jcp135,nazir-prl103,mccutcheon-jcp135,kolli-jcp135}
can potentially resolve such issues.

A minimal model for dissipative quantum processes that proceed through multiple states 
consists of donor-bridge-acceptor
($D$-$B$-$A$) states coupled to a phonon bath.  In the limit where all of the electronic
couplings are small and the energy of state $B$ is comparable with those of states $D$ and $A$,
the dynamics may be described in terms of hopping between different states,
with rate kernels determined from F\"{o}rster-type theory.  However, if the energy of
state $B$ is substantially different from the energies of states $D$ and $A$, such
that thermal activation is infeasible, then a
quantum mechanical super-exchange mechanism must be invoked. 
Super-exchange involves rates that are
fourth-order in the electronic coupling, and thus can be handled by Redfield-like
approaches (which treat the electronic coupling non perturbatively). 
This interplay between sequential hopping and super-exchange has a rich history
in the mechanistic understanding of the primary electron transfer event in
photosynthesis~\cite{hu-cpl160,parson-biochimica1017,marchi-jacs115,makri-pnas96}.

Despite many analyses of experimental data, understanding how and when the crossover from
hopping to super-exchange occurs is not well understood.  It is often suggested in simple
rate theories~\cite{adv-et,kornyshev-pnas103,davis-nature396,vura-weis-science328} that the
sum of the two rates, $k=k_{\rm hop}+k_{\rm SE}$, should  serve as a reasonable approximation.
This implicitly assumes that the two mechanisms can be viewed as additively independent
processes.   This assumption has been questioned on the grounds that hopping and
super-exchange should be considered as two different limits of the same quantum dynamical
process~\cite{saito-jcp131}.  A Pad\'{e}-resummed rate expression for multi-site systems,
first derived by Mukamel and co-workers~\cite{sparpaglione-jcp88,hu-jcp91},
presents one example of a unified formalism
capable of describing both process.  Furthermore, many biological or organic energy conversion
systems correspond to an intermediate coupling situation where the electronic,
electron-phonon, and thermal energy scales are all comparable~\cite{cheng-arpc60}.
In such cases, even the
validity of perturbative rate approaches can be called into question.  

In the present work, we examine the real-time population dynamics of $D$-$B$-$A$ systems
in detail by comparing a recently developed PQME approach with approaches that treat to
second order either the electron-phonon coupling (Redfield theory) or the electronic
coupling (a Marcus or F\"{o}rster-type theory that will henceforth be called the 
``hopping'' approach).  As will be detailed below, the PQME-based theory is not a simple
perturbation theory and the expansion parameter contains information about both the
electron-phonon coupling and the electronic coupling.  In other words, the theory captures
nonperturbative effects in both the electronic and system-bath couplings. 
Our goal is to demonstrate that the
PQME approach correctly interpolates between the two limits described above, making it a
promising candidate for a theory that is valid in all parameter regimes.   

An alternative approach is the so-called modified Redfield theory, which
also interpolates between the weak-coupling Redfield and strong-coupling F\"{o}rster
theories~\cite{zhang-jcp108,yang-cp282},
however there are two important differences.  First, the interpolation
is tuned by the electronic energy gap and not the strength of the system-bath coupling,
such that the limiting behavior is not always the physically correct one
(for example, the theory always reduces to Redfield theory in the case of degenerate
energy levels, and is accordingly confined to weak system-bath coupling for this case).
Second, the approach yields a rate description of population dynamics in the basis
that diagonalizes the system Hamiltonian (the exciton basis in the context of photosynthetic
energy transfer), and not in the original basis of the problem (the site basis
in the same context).  This approach might therefore yield a dynamical observable different
than the one of interest.

Previous studies have frequently considered the rate behavior of two-level quantum
systems, using for example modified Redfield theory~\cite{yang-cp282} or more recently
a Markovian version of the PQME approach employed here~\cite{chang-jcp137}.
In light
of the recent interest in the role of quantum coherence in biological systems,
we emphasize that the three-level systems considered here are 
much more enlightening than two-level systems
with regards to the contribution of coherent transport effects.
The effective tunneling through energetic bridge states barriers as embodied in
super-exchange phenomena is a true coherent effect which is straightforwardly
identified and characterized.  The contribution of this effect to the overall rate,
as compared to an activated hopping process can be quantified.  These simple
metrics are much more illuminating than the non-rigorous but 
frequently used ones in two-level systems, such as the timescale over which population
oscillations are observed.  

While a large number of exact numerical studies have been devoted to models with
Ohmic baths~\cite{makri-pnas96,ishizaki-pnas106},
relatively little is known for the super-Ohmic case, which is a more appropriate model for
phonons in many contexts, such as excitation transport in crystals.  
The PQME approach is particularly useful in
this important case, and we will consider only super-Ohmic coupling in this work.  This
fact, however, means that we are unable to provide exact benchmark results for comparison.
Regardless, we will show that the PQME approach naturally recovers the correct
behavior in the two limits where either the system-bath or electronic couplings may be
treated up to second order.   In this sense, we will demonstrate that the PQME approach serves
as a viable approach for the
intermediate regime where neither a hopping nor a super-exchange description is
appropriate.  

This paper is organized as follows. In Section II we describe the model and introduce
parameters relevant to organic and biological charge and energy transfer systems.
Section III provides a brief description of the PQME approach as well as the perturbative
hopping and Redfield theories to which we compare.  Section IV provides results 
and analysis of our
model calculations and we conclude in Section V.

\section{Model}

Consider a $D$-$B$-$A$ system linearly coupled to a bosonic bath.  Let us
denote $D$, $B$, and $A$ as $1$, $2$, and $3$, respectively. The total Hamiltonian is
given by  $H=H_{\rm s}+H_{\rm b}+H_{\rm sb}$, where the system Hamiltonian is
\begin{equation}
H_{\rm s}=E_1|1\rangle\langle 1|+E_2|2\rangle\langle 2|+E_3|3\rangle
\langle 3| +J_{12}(|1\rangle\langle 2|+|2\rangle\langle
1|)+J_{23}(|2\rangle\langle 3|+|3\rangle\langle 2|),
\end{equation}
In the above
expression, $E_l$ is the site (or excitation) energy of state $|l\rangle$.  $J_{ll'}$ is
the electronic coupling between states $|l\rangle$ and $|l'\rangle$.
We emphasize that donor and acceptor states 1 and 3 are not directly
coupled.
Therefore any population transfer from donor to acceptor must be mediated, either
physically or virtually, by the bridge state 2.
The bath Hamiltonian
is given by $H_{\rm b}=\sum_n\hbar \omega_n (b_n^\dagger b_n+\frac{1}{2})$ and the
system-bath interaction has the form $H_{\rm sb}=\sum_{l=1}^3\sum_n\hbar \omega_n
(b_n+b_n^\dagger) g_{n,l}|l\rangle\langle l| $.  Each site is assumed to be coupled to an
independent super-Ohmic bath, although consideration of common modes or much more general
and
correlated spectral densities is possible~\cite{jang-jcp129,jang-jcp131,jang-jcp135}.
The spectral density for each state $l$ is given by the super-Ohmic form
\begin{equation}
{\mathcal J}_l(\omega)\equiv\sum_n \delta (\omega-\omega_n) \omega_n^2 g_{n,l}^2 \
=\frac{\eta_l}{3!}\frac{\omega^3}{\omega_{c,l}^2} \rm\rme^{-\omega/\omega_{c,l}} \ .
\end{equation}
We define the corresponding bath correlation function as
\begin{equation}
{\mathcal C}_l(t)=\int_0^\infty d\omega \frac{{\mathcal J}_l(\omega)}{\omega^2}
    \left [\coth(\hbar\omega/2k_B T)\cos (\omega t)-\rmi\sin (\omega t)\right ],
\end{equation}
which fully characterizes the linear response of the bath.   The second
derivative of the above time correlation function is needed in Redfield
theory,
\begin{equation}
{\mathcal C}^{(2)}_l(t)=\int_0^\infty d\omega {\mathcal
J}_l(\omega) \left [ \coth(\hbar\omega/2k_B T)\cos (\omega t)-\rmi\sin (\omega
t)\right ].
\end{equation}
It is interesting to note that ${\mathcal C}_l(t)$ with a
super-Ohmic spectral density is equivalent to ${\mathcal C}_l^{(2)}(t)$ with an
Ohmic spectral density up to a constant factor.   This may explain why the
use of Ohmic spectral densities in Redfield-like QMEs often appears to reproduce
experimental data even when the actual spectral density might be closer to the
super-Ohmic case. 

\section{Methods}

\subsection{Polaronic QME}

The main theoretical tool used for exploring
$D$-$B$-$A$ dynamics in this work is a recently developed PQME approach that combines the
conventional QME approach with a polaron
transformation~\cite{jang-jcp129,jang-jcp131,jang-jcp135}. A short review of this method is
provided first. 

Given the total density operator $\rho(t)$, one can introduce a 
polaron-transformed density operator, $\tilde \rho(t)=\rme^G \rho \rme^{-G}$, where
$G=\sum_{l} \sum_n g_{n,l} (b_n^\dagger-b_n) |l\rangle\langle l|$.   Since $\rme^G$ is
unitary in the total space of system and bath states, any physical observable can be
calculated by taking the trace of $\tilde \rho (t)$ with the corresponding transformation of
the physical observable, regardless of whether an actual polaron is formed or not.   The time
evolution of $\tilde \rho (t)$ is governed by the quantum Liouville equation corresponding
to the polaron-transformed Hamiltonian
\begin{equation}
\tilde H=\rme^GH\rme^{-G}=\sum_{l=1}^N \tilde E_l|l\rangle\langle l|+\sum_{l\neq l'}^N
J_{ll'}\theta_{l}^\dagger \theta_{l'} |l\rangle\langle l'|+H_{\rm b}\ , \label{eq:tilde_h}
\end{equation}
where $\tilde E_l=E_l-\sum_n \hbar\omega_n g_{n,l}^2$ and
$\theta_l=\rme^{-\sum_n g_{n,l} (b_n^\dagger-b_n)}$.  
The state energies are thus shifted by the reorganization energy
$\lambda_l = \sum_n \hbar\omega_n g_{n,l}^2$ and the off-diagonal couplings
have acquired a dynamical modulation due to the bath
degrees of freedom.  Taking the thermal average of these dynamical couplings, we define
the zeroth-order Hamiltonian as
\begin{equation}
\tilde H_0 = \sum_l \tilde E_l |l\rangle\langle l|+\sum_{l\neq
l'}J_{ll'}w_{ll'}|l\rangle\langle l'|+H_{\rm b} = \tilde H_{0,\rm
s}+H_{\rm b}\ , 
\end{equation}
where $w_{ll'}=\Tr_{\rm b}\{
\theta_l\theta_{l'}\rho_{\rm b}\}=\rme^{-\sum_n \coth (\hbar \omega_n/2k_B T) \delta
g_{n,ll'}^2/2}$ with $\delta g_{n,ll'}=g_{n,l}-g_{n,l'}$,
such that the perturbation Hamiltonian,
defined as the difference between $\tilde H$ and $\tilde H_0$, is given by 
\begin{equation}
\tilde H_1=\sum_{l\neq l'}J_{ll'}\left\{\theta_l^\dagger
\theta_{l'}-w_{ll'}\right\} |l\rangle \langle l'|\ .
\label{eq:til_h1}
\end{equation}
In the interaction picture with respect to $\tilde H_0$, the corresponding density
operator, $\tilde \rho_I(t)=\rme^{\rmi\tilde H_0 t/\hbar}\tilde \rho (t) \rme^{-\rmi\tilde H_0
t/\hbar}$, evolves according to the time-dependent Liouville equation
\begin{equation}
\frac{\rmd}{\rmd t}\tilde \rho_I(t)=-\frac{\rmi}{\hbar}\left [\tilde H_1(t), \tilde
\rho_I(t)\right ]  \  \label{eq:til_rho_i} \end{equation}
where
\begin{equation}
\tilde
H_1(t)=\sum_{l\neq l'}J_{ll'}\left\{\theta_l^\dagger
(t)\theta_{l'}(t)-w_{ll'}\right\}{\mathcal T}_{ll'}(t)\ , \label{eq:til_h1_t}
\end{equation}
with $\theta_l(t)=\rme^{-\sum_n g_{n,l} (b_n^\dagger \rme^{\rmi\omega_n
t}-b_n \rme^{-\rmi\omega_n t})}$ and ${\mathcal T}_{ll'}(t)=\rme^{\rmi\tilde H_{0,\rm s}
t/\hbar} |l\rangle\langle l'| \rme^{-\rmi\tilde H_{0,\rm s} t/\hbar}$.

Therefore, the parameter to be treated perturbatively in the PQME approach
is the instantaneous \textit{fluctuation}
of the bath-modulated hopping from its average value.
It is important to note that the
re normalized system-bath coupling $\tilde H_1(t)$ vanishes in the limit of weak
system-bath coupling,
$\eta_l\rightarrow 0$, and remains bounded by $J_{ll'}$ in the strong coupling limit,
$\eta_l\rightarrow \infty$.  If $J_{ll'}\sqrt{1-w_{ll'}}$ is sufficiently small,
truncating the formally exact QME at the second-order of $\tilde H_1(t)$ is valid for all
values of $\eta_l$, and the resulting second order polaronic QME can serve as a good
approximation covering the entire regime of system-bath coupling. 

Using
projection operator techniques~\cite{vankampen-jsp87,jang-jcp116}, one can derive a 
time-local equation of motion for the reduced density operator,
$\tilde \sigma_I(t) = \Tr_{\rm b}\{\tilde \rho_I(t)\}$,
yielding
\begin{equation}
\frac{\rmd \tilde \sigma_I(t)}{\rmd t}
    =-{\mathcal R}(t)\tilde \sigma_I(t)+{\mathcal I} (t) \ , \label{eq:sigma_it}
\end{equation}
where
\begin{eqnarray}
{\mathcal R}(t)\tilde \sigma_I(t)&=\frac{1}{\hbar^2}\sum_{l\neq
l'}\sum_{m \neq m'} J_{ll'} J_{mm'} w_{ll'} w_{mm'} \nonumber \\
&\hspace{1em} \times \int_0^t  d\tau
\left( \rme^{-{\mathcal K}_{ll',mm'} (t-\tau)}-1 \right )[{\mathcal T}_{ll'}(t),{\mathcal
T}_{mm'}(\tau) \tilde \sigma_I (t)] +{\rm H.c.} \ ,
\end{eqnarray}
with ${\mathcal K}_{ll',mm'}(t)=(\delta_{lm}-\delta_{lm'}){\mathcal
C}_{l}(t)+(\delta_{l'm'}-\delta_{l'm}){\mathcal C}_{l'}(t)$.  The source term 
${\mathcal I}(t)$ arises for generically nonequilibrium initial conditions.
Explicit expressions for this term can be found for the general case where
the initial excitation is an arbitrary superposition of the system
states~\cite{jang-jcp135}. The reduced equation of motion 
can be solved numerically in the eigenbasis of $\tilde H_{0,s}$ as
detailed previously~\cite{jang-jcp131}.  Unlike $\tilde \rho_I(t)$, not all the system
observables can be calculated from $\tilde \sigma_I(t)$, but because
$\rme^G |l\rangle \langle l| \rme^{-G} = |l\rangle \langle l|$, determination of the site
populations is still possible via
\begin{equation}
P_l(t)=\Tr_{\rm s}\left\{ \rme^{\rmi\tilde
H_{0,\rm s}t/\hbar} |l\rangle\langle l|\rme^{-\rmi\tilde H_{0,\rm s}t/\hbar} \tilde
\sigma_I(t)\right\}\ ,
\end{equation}
where $\Tr_{\rm s}$ denotes trace over the system degrees
of freedom. 

\subsection{Hopping dynamics}
The master equation that we henceforth refer to as embodying hopping dynamics can
also be derived via
projection operator techniques, expanding directly in the bare electronic coupling
elements $J_{lm}$.  The resulting equation of motion for the populations terms is
closed and given by the Pauli master equation
\begin{equation}
\frac{\rmd P_l(t)}{\rmd t} = \sum_{m\neq l}
    \left\{ k_{m\rightarrow l}^{F}(t) P_m(t) - k^{F}_{l\rightarrow m}(t) P_l(t) \right\}.
\label{eq:pauli_m} 
\end{equation}
For the time-dependent rates
in the above expression, we employ the non-Markovian version of Fermi's Golden
rule~\cite{jang-cp275}, 
\begin{equation}
k^F_{l\rightarrow m}(t)
=\frac{2(J_{lm}w_{lm})^2}{\hbar^2} {\rm Re} \int_0^t d\tau
\rme^{\rmi (\tilde E_l -\tilde E_m)\tau /\hbar} \left ( \rme^{{\mathcal
K}_{lm,lm}(\tau)}-1\right)    \ . 
\end{equation}
Note that a contribution proportional
to $\delta (\tilde E_l-\tilde E_m)$ is subtracted from the above integration.  This
regularization makes the resulting integration convergent for the super-Ohmic spectral
density considered here, which amounts to subtracting the contribution of the zero phonon
line to the rate.  Note that, aside from the initial condition term $\mathcal{I}(t)$,
the main difference between the PQME and this master equation that describes hopping 
dynamics is that the latter treats the bare electronic coupling as a perturbation,
while the PQME treats the \textit{fluctuations} of the electronic coupling terms in
the polaronic basis as a perturbation.
The hopping dynamics can thus only predict quantum effects that are second-order
in the couplings $J_{lm}$.

\subsection{Redfield theory}

At the opposite extreme,
we consider Redfield theory~\cite{red65,bre02},
which is a QME approach that uses the system-bath coupling as a perturbation
and treats the entire system Hamiltonian exactly.
As such, this approach encompasses the conventional fourth-order super-exchange result in the
high barrier limit and furthermore includes all higher-order electronic interactions
for moderate or small barrier heights.

In the interaction picture with respect to $H_{\rm s}$, the Redfield theory QME takes
the form
\begin{equation}
\frac{\rmd \sigma_I(t)}{\rmd t} =- R \sigma_I(t)\ , 
\end{equation}
where
\begin{equation}
R \sigma_I(t) = \frac{-1}{\hbar^2} \sum_{l\neq l'}\sum_{m \neq m'}
    \int_0^\infty  d\tau {\mathcal K}^{(2)}_{ll',mm'} (t-\tau)
    [T_{ll'}(t),T_{mm'}(\tau) \sigma_I (t)] +{\rm H.c.} \ ,
\end{equation}
In the above, ${\mathcal K}^{(2)}_{ll',mm'} (t)$ is the second time derivative of
${\mathcal K}_{ll',mm'} (t)$ defined in the previous section and
$T_{ll'}(t)=\rme^{\rmi H_{\rm s} t/\hbar} |l\rangle\langle l'| 
    \rme^{-\rmi H_{\rm s} t/\hbar}$.
We furthermore employ the secular approximation, where all
elements $R_{jj'}^{kk'}$ in the basis of system eigenstates
for which $|\delta {\mathcal E}_{kk'}^0-\delta {\mathcal
E}_{jj'}^0|\neq 0$ are neglected ($\delta {\mathcal E}_{kk'}^0$ is the difference
between the eigenenergies of system eigenstates $k$ and $k'$).  The secular approximation
prevents unphysical negative or diverging populations in the limit of strong
system-bath coupling, where the second-order approximation inherent in the Redfield
approach breaks down.  More specifically, the secular approximation enforces
the equilibrium population $\sigma^{\rm eq} \propto \rme^{-H_s/k_B T}$ and kinetic rates
$k \propto \eta$ for all values of the system-bath coupling, whereas these results
are clearly only correct for weak coupling.

\section{Results}

To probe generic effects of bridge energetics on the mechanistic and
quantitative details of population transfer from donor to acceptor,
we investigate a small set of model parameters close to those
encountered in light harvesting complexes or singlet fission problems.
In particular, we are
interested in the situation where the electronic couplings, width of spectral densities,
and the energy difference between donor and acceptor are all comparable.  Thus, we use
fixed values of $J_{12}=J_{23}=100\ {\rm cm^{-1}}$ and $E_1-E_3=100\ {\rm cm^{-1}}$
for all the calculations (the latter providing an energetic driving force),
and scan a range of physically relevant values for other parameters. 
Furthermore, we fix the bath of all three sites to have identical spectral densities,
such that $\eta_l = \eta$ and $\hbar \omega_{{\rm c},l} = \hbar \omega_{\rm c}$.

First, we consider the case of $\hbar\omega_{\rm c}=200\ {\rm cm^{-1}}$ at temperature
$T=300\ {K}$, where the spectral range of the bath is comparable to the thermal energy.
Both quantum and multiphonon effects of the bath are important in this case.  Nine
different parameter sets, with $\eta=0.2$, $1$, $5$ and $E_2-E_1=0$, $200$, $800\ {\rm
cm^{-1}}$ are investigated.

Figure 1 shows the calculated time-dependent populations of the acceptor state, $P_3(t)$.   
The population based on the PQME exhibits strongly coherent behavior during early
times and becomes increasingly incoherent with increasing system-bath
coupling or bridge state energy. 
In the weak system-bath coupling limit ($\eta=0.2$), the results of the PQME
agree with those of Redfield theory, but those of the hopping dynamics
differ substantially. In the strong system bath-coupling limit ($\eta=5$), the opposite
situation occurs.  The PQME results agree with those of the hopping
dynamics, whereas the Redfield theory results are substantially different.  For
moderate system-bath coupling ($\eta=1$), the degree of agreement is sensitive to
the value of $E_2-E_1$, with the agreement worsening for increasing bridge energy.
Specifically, 
the hopping dynamics agree with the PQME results only
when $E_2-E_1\lesssim k_BT$, consistent with thermal activation.
For the largest value of bridge energy considered, $E_2-E_1=800\ {\rm cm^{-1}}$,
the PQME dynamics are much faster than those of the hopping approach, but slightly
slower than those of Redfield theory. 

\begin{figure}
\begin{center}
\includegraphics[scale=0.375,clip]{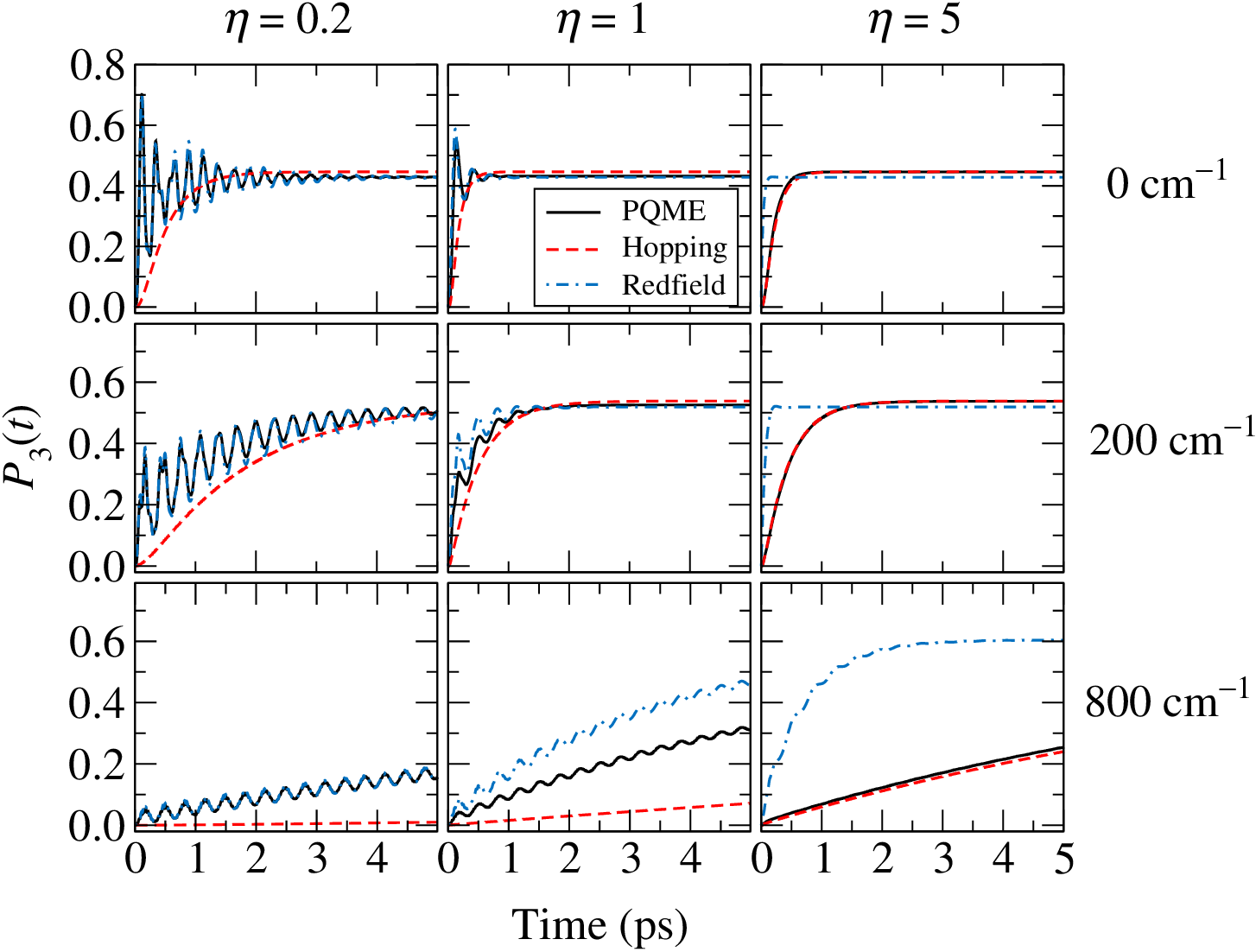}
\end{center}
\caption{Time-dependent populations of the final site $l=3$ based on PQME
(black solid lines), hopping (red dashed lines), and Redfield (blue dot-dashed lines)
theories.
The columns have different values of the system-bath coupling strength
$\eta=0.2$, $1$, and $5$, and rows have different values of the energetic barrier
$E_2-E_1= 0$, $200$, and $800\ {\rm cm^{-1}}$.  Other
parameters are fixed and given by $E_1-E_3= 100\ {\rm cm^{-1}}$, $\hbar\omega_{\rm c}=200\
{\rm cm^{-1}}$, $J_{12}=J_{23}=100\ {\rm cm}^{-1}$, and $T=300$ K.}
\end{figure}

In order to understand the effects of temperature, similar calculations are performed
at a lower
temperature $T=100\ {\rm K}$ and the results are shown in Figure 2.  While similar
trends as in Figure 1 can be seen, the discrepancy between  the results of PQME and hopping
dynamics are much more pronounced for weak and moderate coupling,
and there are still significant differences
between the two even for strong coupling.
In this low temperature limit, quantum effects dominate
not only the dynamics but also the steady state limit.  This suggests that
delocalized exciton states have more physical meaning in this low
temperature limit, for which the localized states (including the system-bath
coupling) are not Boltzmann distributed.

\begin{figure}
\begin{center}
\includegraphics[scale=0.375,clip]{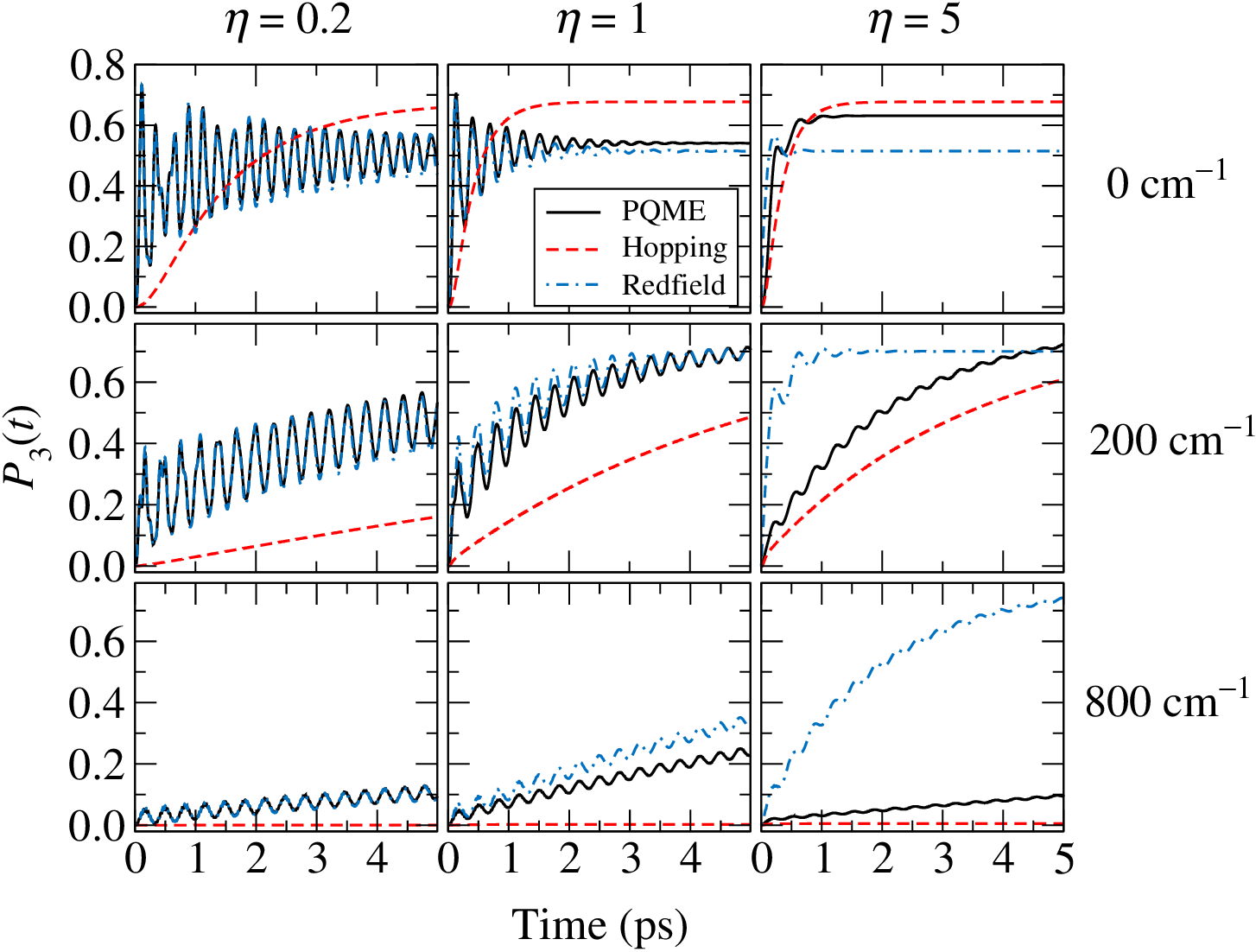}
\end{center}
\caption{
The same as in Figure 1 except for $T=100$ K.
}
\end{figure}

Figure 3 shows analogous results when the spectral density is much
broader than other energy scales, $\hbar \omega_{\rm c} = 1,000\ {\rm cm^{-1}}$.
For each column, the reorganization energy of the bath, which is another measure
of the strength of system-bath coupling
($\lambda \propto \eta \hbar \omega_{\rm c}$), is five times the corresponding
column of Figure 1 or 2.  For example, the case of $\eta=0.2$ in Figure 3
has the same reorganization
energy as the case of $\eta=1$ in Figure 1.
For this situation, the majority of bath modes are no longer resonant with
electronic transitions amongst the system energy levels and thus they do not
effectively induce dephasing.
The resulting dynamics are therefore much more coherent despite the fact that the
reorganization energy is five times larger than in the previous examples.  Clearly
the ability of the PQME approach to accurately capture this
short-time quantum coherence is especially encouraging, given the recent interest
in such phenomena.

\begin{figure}
\begin{center}
\includegraphics[scale=0.375,clip]{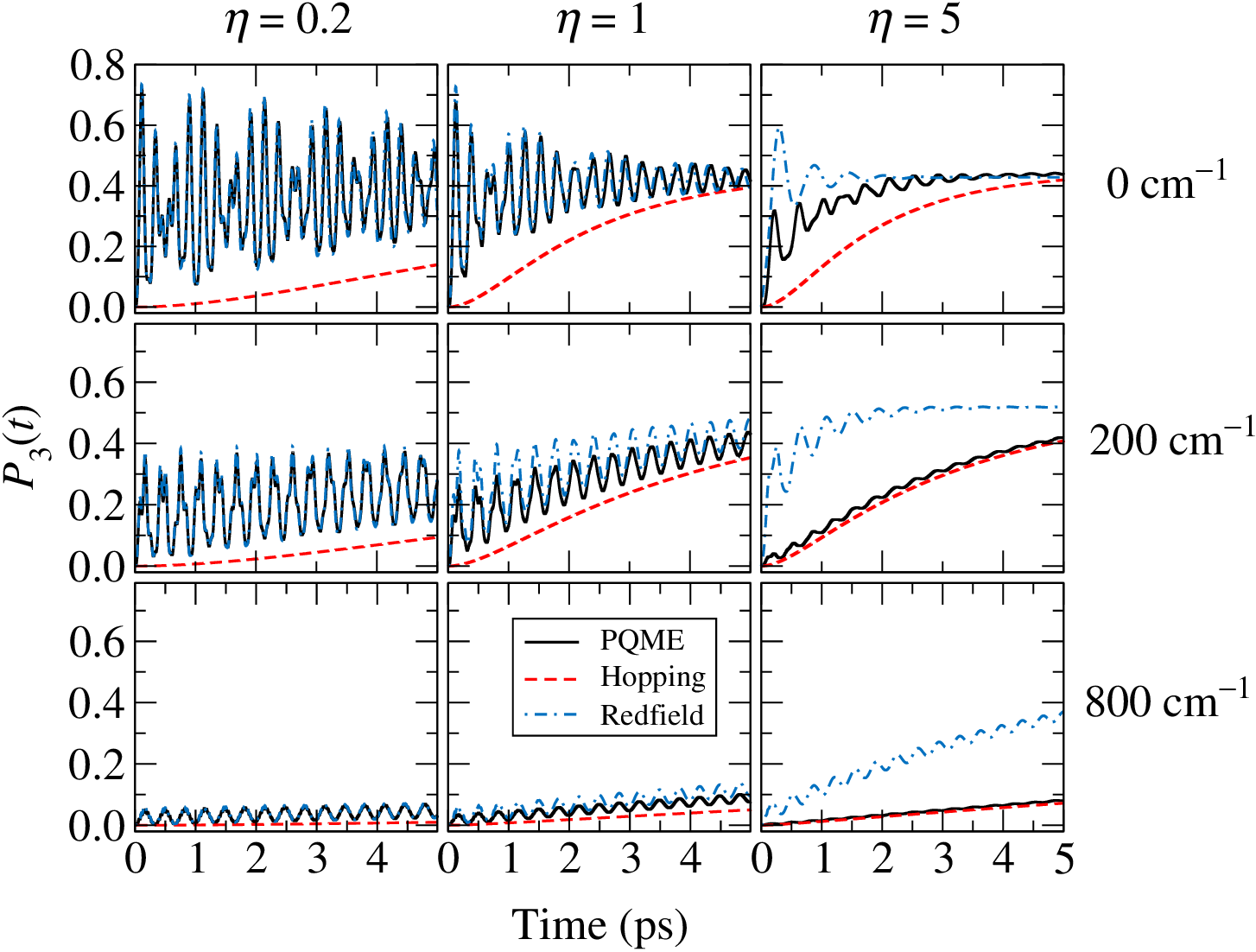}
\end{center}
\caption{
The same as in Figure 1 except for $\hbar \omega_{\rm c}=1,000\ {\rm cm^{-1}}$.
}
\end{figure}

Although many of the results based on the PQME and Redfield theory exhibit 
such coherent
population dynamics, their long-time trends obey nearly exponential decay.  For
the parameters corresponding to the middle column of Figure 1 (i.e.~moderate
system-bath coupling, $\eta=1$), calculations were conducted
for values of $E_2-E_1$ in the range of $100-1,000\ {\rm cm^{-1}}$.  The
resulting time dependent populations were then fit to the kinetic form
\begin{equation}
P_{3}(t)=P_{3}^{\rm eq}(1-\rme^{-k t})\ , \label{eq:rate_fit}
\end{equation}
and the extracted values of $k$ are shown in Figure 4.   The main panel plots the data in
logarithmic scale (base $\rme$), and the inset shows the same data with the energy difference
$E_2-E_1$ in linear scale.  As
shown by the extrapolation expected from super-exchange theory (SE-limit),
the results of the Redfield calculation
clearly demonstrate super-exchange behavior, $k \sim |E_2-E_1|^{-2}$,
for large values of $E_2-E_1$. This
does not necessarily mean that the rates are the same as those calculated by perturbative
(fourth-order) super-exchange theory because there might be non-negligible contributions
from the subleading terms.  On the other hand, the inset clearly demonstrates that the hopping
dynamics exhibit exponential behavior, $k \sim \exp(-|E_2-E_1|/k_B T)$
indicative of barrier crossing via thermal activation.
The results of the PQME calculation instead behave like activated process for small values of
the bridge energy and like super-exchange behavior for large values, where the crossover
is dependent on other factors including the strength of the system-bath coupling.
For all the values of bridge energy considered, the rate predicted by the
PQME theory are in-between those of Redfield theory and hopping dynamics,
suggesting that the simple approximation of
$k=k_{\rm hop}+k_{\rm SE}$ is unreliable.

\begin{figure}
\begin{center}
\includegraphics[scale=0.375,clip]{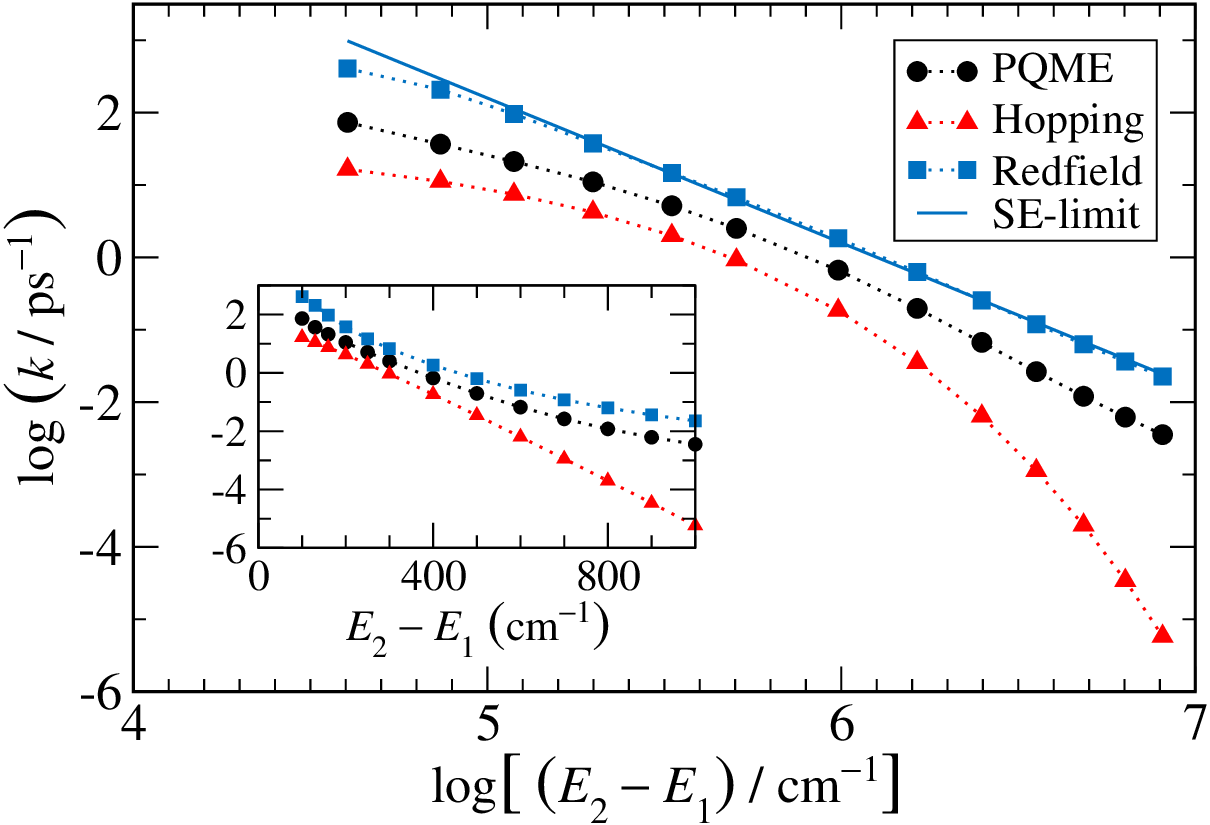}
\end{center}
\caption{Population transfer rates versus $E_2-E_1$ in logarithmic scale (base $\rme$)
for $\eta=1$,
$\hbar \omega_{\rm c}=200\ {\rm cm^{-1}}$, $J_{12}=J_{23}= 100\ {\rm cm^{-1}}$, 
and $T=300\ {\rm K}$. The blue solid line
is the extrapolation of the Redfield results with slope $-2$, demonstrating
super-exchange behavior, $k \sim |E_2-E_1|^{-2}$.
The inset shows the same results with a linear
scale for $E_2-E_1$, demonstrating the
exponential dependence of the hopping rate, $k \sim \exp(-|E_2-E_1|/k_B T)$.}
\end{figure}

However, the above analysis is based on the assumption
that the Redfield rate constant is a reasonable proxy for the super-exchange rate constant,
which is not entirely accurate.  It remains to be determined whether the Redfield
rate is correctly larger than the others because it exactly includes \textit{all} orders
of electronic transport effects or whether the Redfield rate is incorrectly large because
it does not properly include the renormalized hopping (or band narrowing) effect associated
with a finite system-bath coupling.  Without numerically exact data, we cannot answer this
question conclusively, however it is instructive to conclude by considering the dependence
of the rate on the strength of the system-bath coupling.

Interestingly, we can infer from Figure 1 
that the population transfer to the acceptor reaches a maximum for
appropriate values of the system-bath coupling, $\eta$.  In order to explore this aspect
in more detail, we have carried out an analogous study of the rate constant for values
of the system-bath coupling $\eta$ in the range 0.5 -- 10.
The results are shown in Figure 5 for two different
choices of $E_2-E_1=200$ and $800\ {\rm cm^{-1}}$.  It is clear that the Redfield rate
constant is indeed becoming unphysically large with increasing system-bath coupling,
due to the aforementioned weak-coupling prediction, $k \sim \eta$.  On the other hand,
the purely hopping behavior grossly underestimates the rate at small values of the coupling,
where quantum coherent transfer process dominate because the bath cannot effectively
activate classical barrier crossing.  Ultimately, we again see encouraging evidence that
the PQME approach interpolates between the two limiting cases where each theory is most
accurate, and generically exhibits a characteristic turnover behavior, where the turnover
can be ascribed to a self-trapping effect (i.e.~polaron formation in the case of charged
quantum states).  The case of $E_2-E_1 = 800\ {\rm cm^{-1}}$ is especially interesting,
because while Redfield theory and the hopping process predict a rate that
varies by four orders of magnitude over the range studied, the interpolating behavior of the PQME
approach instead predicts a rate which is largely insensitive to the strength of the coupling,
varying by less than one order of magnitude.

\begin{figure}
\begin{center}
\includegraphics[scale=0.375,clip]{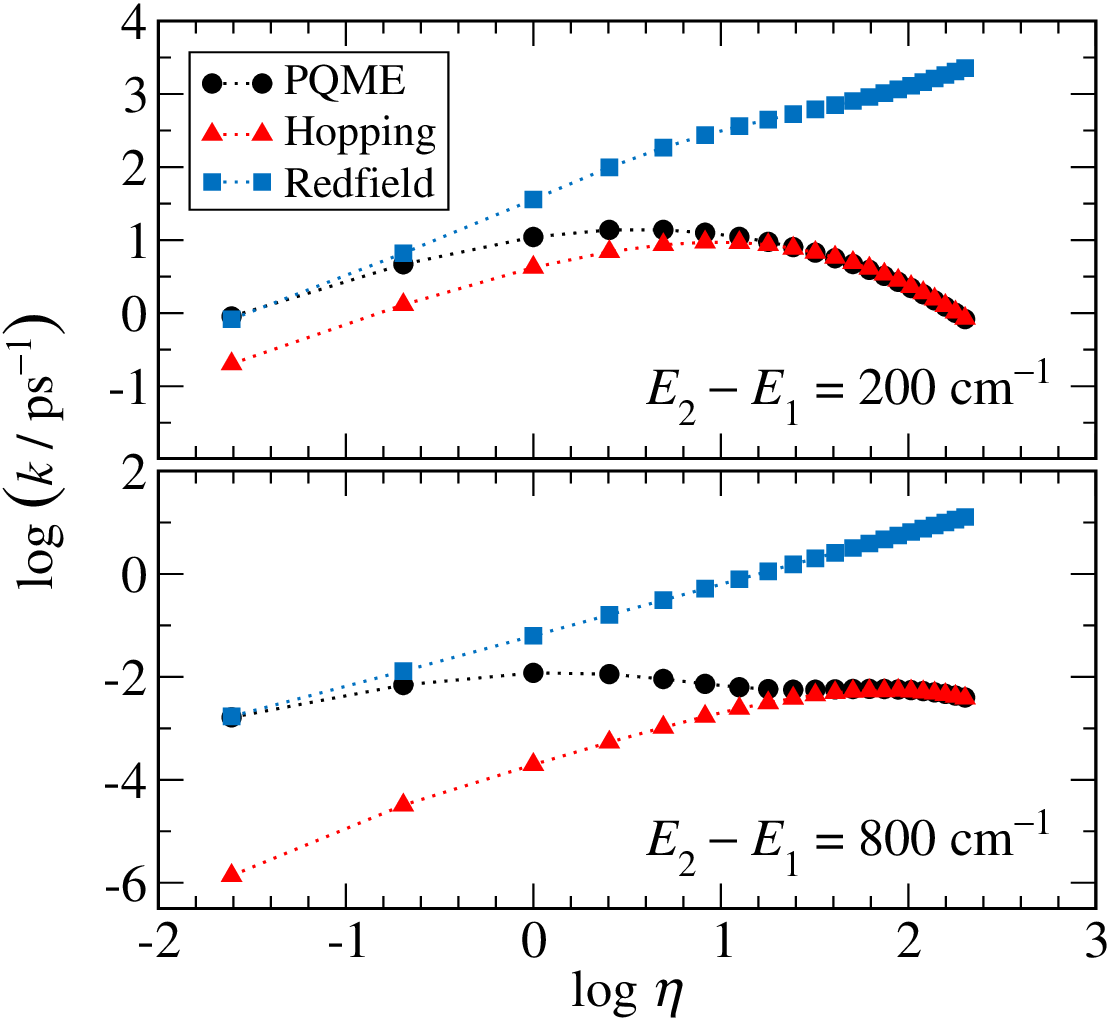}
\end{center}
\caption{Logarithmic dependence (base $\rme$) of population transfer rates on the strength of
the system-bath coupling $\eta$ for $\hbar
\omega_{\rm c}=200\ {\rm cm^{-1}}$, $J_{12}=J_{23}= 100\ {\rm cm^{-1}}$,  and $T=300\ {\rm
K}$.  The upper panel is for $E_2-E_1=200\ {\rm cm^{-1}}$ and the lower panel is for
$E_2-E_1=800\ {\rm cm^{-1}}$. } 
\end{figure}

\section{Conclusion}

In this paper we have investigated the dynamics of a donor-bridge-acceptor system in
detail.  The main goal of the work was to show that the PQME approach can capture
the limiting behaviors of the Redfield and hopping
approaches in the regimes where each is expected to be valid.  These regimes include
both super-exchange kinetics (when the bridge state energy lies energetically well above
the donor and acceptor), as well as incoherent hopping kinetics (when the bridge state
energy is sufficiently low that thermal activation is possible).  The PQME approach
captures these limiting forms by combining a polaron transformation with a novel
definition of the perturbation as a \textit{fluctuation} of the transformed
hopping term.  This protocol yields a perturbative parameter which is a complex mixture of
both system and bath degrees of freedom.  When the system-bath coupling is weak, the
renormalized hopping parameter is close to the bare one and a linearized coupling due to
the fluctuating hopping perturbation is similar to the bare system-bath coupling.  Thus,
the theory approaches the Redfield limit.  When the system-bath coupling is strong, the
renormalized hopping term is vanishingly small and the difference between the fluctuating
and bare hopping parameters is negligible.  Hence, the PQME approach reduces
to a hopping-type
F\"{o}rster theory.  While it will be important to compare the PQME approach to exact
calculations in the future, the fact that it captures the crossover between these two
important and distinct regimes supports the possibility that it also captures the
essential features of the difficult ``intermediate coupling'' regime where most
perturbative approaches fail.  This observation, as well as the efficiency and scalability
of the PQME approach, make it promising for studies of the charge and energy transport in
large systems within parameter regimes that are problematic for standard approaches.

In addition to the comparison between the PQME and other approaches, our study has also
revealed several other interesting facets of the behavior of $D$-$B$-$A$ systems.  In
particular, even in cases where the population exhibits a simple rate behavior, the PQME
dynamics, which include quantum coherence effects, are significantly faster than those of
the second-order hopping process.  Thus, population transfer through a partially
coherent mechanism is clearly significant and becomes more dominant as the bridge
energy increases into the super-exchange regime.  In addition to electronic effects,
we have also highlighted novel, non-perturbative system-bath coupling effects
beyond the reach of other
treatments, such as the robust prediction of a non-monotonic dependence of the
rate on the system-bath coupling strength.  All of these subtle aspects of $D$-$B$-$A$
dynamics are worthy of future study.

%\acknowledgments
\ack

This research was conducted during a sabbatical stay of SJ at Columbia University.
The research contributions of SJ, TCB, and DRR were supported primarily by
the Center for Re-Defining Photovoltaic Efficiency Through
Molecule Scale Control, an Energy Frontier Research Center funded by the U.S. Department
of Energy, Office of Science, Basic Energy Sciences under Award No.~DE-SC0001085.
SJ also
acknowledges the partial support of the National Science Foundation CAREER award (Grant No.
CHE-0846899), the Office of Basic Energy Sciences, Department of Energy (Grant No.
DE-SC0001393) for preliminary development and calculation conducted at Queens College.
SJ was partially supported through a Camille Dreyfus Teacher Scholar Award.
TCB was partially supported by the Department of Energy Office of Science Graduate Fellowship
Program (DOE SCGF), administered by ORISE-ORAU under Contract No.~DE-AC05-06OR23100.

\end{document}